\documentclass[%
 reprint,
 amsmath,amssymb,
]{revtex4-2}

\usepackage{graphicx}
\usepackage{dcolumn}
\usepackage{bm}
\usepackage{ulem}

\usepackage{amssymb} 
\usepackage{multirow}
\usepackage{booktabs}
\usepackage{color}     
\usepackage{amsmath}
\usepackage{multirow}  

\usepackage{hyperref} 
\hypersetup{
            colorlinks=true,
            linkcolor=blue,
            filecolor=magenta,      
            urlcolor=cyan, 
            citecolor=red,}
            
\begin{document}

\preprint{APS/123-QED}

\title{Nuclear matrix element of $2\nu\beta\beta$ decay of $^{76}$Ge: roles of high-lying states and two-body currents } 

\author{Hua-Yang Xu}%
\affiliation{School of Physical Science and Technology, Southwest University, Chongqing 400715, China}%

\author{Hao Zhou}%
\affiliation{School of Physical Science and Technology, Southwest University, Chongqing 400715, China}%

\author{Long-Jun Wang}
\email{longjun@swu.edu.cn}
\affiliation{School of Physical Science and Technology, Southwest University, Chongqing 400715, China} 

\date{\today}

\begin{abstract}
We present a microscopic analysis of the nuclear matrix element (NME) of the two-neutrino double-$\beta$ ($2\nu\beta\beta$) decay for open-shell heavy deformed nuclei, taking into account the fact that nuclear level density increases rapidly with excitation energy as well as the contribution of two-body current (2BC). Taking $^{76}$Ge $\rightarrow$ $^{76}$Se decay as an example, we found that due to the rapid increase of the level density of the intermediate nucleus $^{76}$As with excitation energy $E_n$, the single-$\beta$ Gamow-Teller (GT) matrix elements become highly fragmented with very small magnitude, and exhibit seemingly random sign patterns at high $E_n$ region. This leads to an effective cancellation at high $E_n$ region in calculating the $2\nu\beta\beta$ NME which then turns out to converge at $E_n \lesssim 5$ MeV, indicating that the contribution of high-lying states of the intermediate nucleus to $2\nu\beta\beta$ NME is negligible. Besides, the 2BC in the transition operator is found to contribute $\sim 10\%$ quenching to the $2\nu\beta\beta$-decay NME of $^{76}$Ge.
\end{abstract}

\maketitle


\section{\label{sec:intro}Introduction }

The search for neutrinoless double-$\beta$ ($0\nu\beta\beta$) decay represents a major frontier in probing physics beyond the Standard Model. As a lepton-number-violating process, its observation would unambiguously confirm the Majorana nature of neutrinos and provide access to the absolute neutrino mass scale \cite{Haxton_PPNP_1984, Doi_1985, Tomoda_RPP_1991, Suhonen_Phys_Rep_1998, Avignone_RMP_2008, Vergados_2012_RPP, Engel_2017_RPP, bilenky_2018_book, Suhonen_Phys_Rep_2019, JMYao_PPNP_2022}. Among various candidates, $^{76}$Ge has attracted extensive experimental attention, including GERDA, MAJORANA DEMONSTRATOR, the next-generation tonne-scale LEGEND \cite{Agostini_PRL_2020, Arnquist_PRL_2023, Burlac_2025} etc. 

The nuclear matrix element (NME) is the key to understand the related physics of $0\nu\beta\beta$ decay. However, the NME of $0\nu\beta\beta$ decay is not an observable and it is challenging to provide a reliable theoretical description. On the other hand, as the counterpart, the two-neutrino double-$\beta$ ($2\nu\beta\beta$) decay is allowed by the Standard Model and has been measured experimentally. The extracted NME of $2\nu\beta\beta$ decay can provide helpful constraints on nuclear-structure models that are applied to calculate the NME of $0\nu\beta\beta$ decay, since both decay modes share the identical initial and final nuclear states. 

Except for the initial and final states of parent and daughter nuclei, the description of the NME of $\beta\beta$ decay (which is a second-order weak-interaction process) also involves two aspects, one is the nuclear states, including low-lying and high-lying ones, of the intermediate nucleus, the other is the weak axial-vector coupling constant $g_A$ in nuclear medium which may include possible quenching mechanism \cite{Javier2011PRL, LJWang_current_2018_Rapid}. For the former, nuclear level density increases very rapidly with excitation energy \cite{Guttormsen_Level_Density_EpJA_2015, JQWang_PRC_2023}, especially for open-shell nuclei which are relevant to most $\beta\beta$-decay candidates. This means, there exist thousands of virtual single-$\beta$ transitions that connect the thousands of levels within each 1 MeV bin of the intermediate nucleus with the initial (final) state of the parent (daughter) nucleus. Each matrix element of the virtual single-$\beta$ transitions should be very small with different phases (signs), summing them up yields the total $\beta\beta$-decay NME. This indicates that cancellation may occur in the summation \cite{Brown_PRC_1990, Ericson_PLB_1994}, which is important in understanding the $\beta\beta$-decay NME. On the other hand, the latter involves the issue of $g_A$ quenching. One probably needs to go beyond the impulse approximation and consider the contributions from two-body current (2BC) in the transition operators, without the closure approximation and normal ordering approximation. 
 
Over the past few decades, great efforts have been made to extract, constrain or calculate the $\beta\beta$-decay NME experimentally and theoretically \cite{Simkovic_PRC_2009, Horoi_PRL_2013, BABrown_PRC_2015, Iachello_PRC_2015, JMYao_2018_PRC, Coraggio_PRC_2019, ECEC_Nature_2019, YKWang_Sci_Bull_2024, Yao_PRL_2024, Satula_PRC_2025_Editor, Vincenzo_JHEP_2025, Thies_76Ge_PRC_2012}. Experimentally, nuclear charge-exchange reactions can provide insight into and constraints on the $\beta\beta$-decay NME. However, charge-exchange reactions only detect the Gamow-Teller (GT) part, which is relevant to $2\nu\beta\beta$ decay, and only measure the GT strengths, which lack the sign information that is crucial for deriving the $\beta\beta$-decay NME. Theoretically, it is still challenging to treat the high level density of the intermediate nucleus properly and consider the two-body currents in the transition operators without the closure approximation and the normal ordering approximation. In this paper, we adopt a shell-model diagonalization method with large model space and large configuration space with the help of angular-momentum projection and the Pfaffian algorithm \cite{LJWang_2014_PRC_Rapid, LJWang_2018_PRC_GT, ZRChen_2022_PRC, BLWang_1stF_2024}, to calculate the $2\nu\beta\beta$-decay NME of a typical open-shell deformed nucleus $^{76}$Ge. We consider both the high level density of the intermediate nucleus and the two-body currents without the closure approximation and/or the normal ordering approximation for the first time. The similar work for the counterpart of $0\nu\beta\beta$-decay NME will be accomplished in the near future. 

The paper is organized as follows. In Sec. \ref{sec:theory} we show the theoretical framework for the calculation of the $2\nu\beta\beta$-decay half-life and NME, including the transition operators and nuclear many-body wave functions. The numerical analysis of the $2\nu\beta\beta$-decay half-life and NME for $^{76}$Ge are shown in Sec. \ref{sec:result}, and a brief summary is given in Sec. \ref{sec:summary}.

\section{\label{sec:theory} Theoretical Framework }

The half-life of $2\nu\beta\beta$ decay is expressed as,
\begin{eqnarray} \label{eq.half_life}
  (T_{1/2}^{2\nu})^{-1} = G^{2\nu} | \mathcal M^{2\nu} |^2 ,
\end{eqnarray}
where $G^{2\nu}$ is the phase space factor which is adopted as $G^{2\nu} = 0.4817 \times 10^{-19} \text{ yr}^{-1}$ here for the $2\nu\beta\beta$ decay of $^{76}$Ge \cite{Kotila_PRC_2012_phase_space}, and $\mathcal M^{2\nu}$ is the (dimensionless) $2\nu\beta\beta$-decay NME which is calculated by explicitly summing over single-$\beta$ matrix elements connecting all intermediate $1^+_n$ levels, 
\begin{eqnarray}\label{eq.double_NME}
  \mathcal M^{2\nu} = m_e \sum_{n} \frac{M_n}{ E_n + E_0 },
\end{eqnarray}
where $M_n = \int d \bm{x}_1 d \bm{x}_2 \langle 0_f^+ \| \hat{\mathcal{J}}(\bm{x}_2) \| 1_n^+ \rangle \langle 1_n^+ \| \hat{\mathcal{J}}(\bm{x}_1) \| 0_i^+ \rangle$. Here $E_n$ is the excitation energy of the $1_n^+$ level of $^{76}$As, and $E_0 = \frac{1}{2} Q_{\beta\beta}(0^+) + \Delta M = 1.94103$ MeV where $Q_{\beta\beta}(0^+) = 2.03906$ MeV corresponds to the $\beta\beta$-decay $Q$ value of $^{76}$Ge and $\Delta M$ is the $^{76}$As -$^{76}$Ge mass difference ($0.9215$ MeV) \cite{Wang_2021_CPC_AME2020}. 

To obtain the $2\nu\beta\beta$-decay NME, the single-$\beta$ matrix elements $\int d \bm{x}_2 \langle 0_f^+ \| \hat{\mathcal{J}}(\bm{x}_2) \| 1_n^+ \rangle$ and $\int d \bm{x}_1 \langle 1_n^+ \| \hat{\mathcal{J}}(\bm{x}_1) \| 0_i^+ \rangle$ connecting the initial ($i$) and final ($f$) $0^+$ state of the parent and daughter nucleus, respectively, with the thousands of levels within each 1 MeV bin of the intermediate nucleus are indispensable. Here the transition operator corresponds to the nuclear weak currents which, when considering only the axial currents in the limit of zero momentum transfer, include the one-body current (1BC) term, two-body current term etc. The 1BC is written in the first-quantized form as \cite{LJWang_current_2018_Rapid, Jon_Engel_currents_PRC_2022, Park_2003_PRC},
\begin{eqnarray} \label{eq.one_body_current}
  \hat{\bm{\mathcal J}}_{1b}(\bm x) = - g_A \sum_{i=1}^A \bm\sigma_i \tau^-_i \delta(\bm x - \bm r_i), 
\end{eqnarray}
where $g_A \approx 1.27$ is the weak axial-vector coupling constant \cite{Axial_Vector_PRL_2019}, $\bm\sigma$ denotes the Pauli spin operator, and $\tau^-$ is the isospin lowering operator with the convention $\tau^-|n\rangle = |p\rangle$. The 1BC corresponds to the GT operator. 

To further consider the 2BC, we follow \cite{Park_2003_PRC} and neglect the term with coefficient $c_6$. The leading space piece of the axial 2BC operator in coordinate space is, \cite{LJWang_current_2018_Rapid, Jon_Engel_currents_PRC_2022},
\begin{widetext}
\begin{eqnarray} \label{eq.two_body_current}
  \hat{\bm{\mathcal J}}_{2b}(\bm x) &=& \sum_{k<l}^A \bm J_{kl}(\bm x), \nonumber \\ 
  \bm J_{kl}(\bm x) &=& \frac{2 \bar c_3 g_A}{m_N F^2_\pi} \left\{   m^2_\pi\left[\left( \frac{\bm\sigma_l}{3} - \bm\sigma_l \cdot \hat{\bm r} \hat{\bm r} \right) Y_2(r) - \frac{\bm\sigma_l}{3} Y_0(r) \right] + \frac{\bm\sigma_l}{3} \delta(\bm r)  \right\} \tau^-_l \delta(\bm x - \bm r_k) + (k \leftrightarrow l) \nonumber \\
  && + \left( \bar c_4 + \frac{1}{4}\right) \frac{g_A}{2 m_N F_\pi^2} \left\{ m_\pi^2 \left[ \left( \frac{\bm\sigma_\times}{3} - \bm\sigma_k \times \hat{\bm r} \bm\sigma_l \cdot \hat{\bm r} \right) Y_2(r) - \frac{\bm\sigma_\times}{3} Y_0(r) \right] + \frac{\bm\sigma_\times}{3} \delta(\bm r) \right\} \tau^-_\times \delta(\bm x - \bm r_k) + (k \leftrightarrow l) \nonumber \\
  && - \frac{g_A}{4m_N F_\pi^2} \left[ 2\bar d_1 \left( \bm \sigma_k \tau^-_k + \bm\sigma_l \tau^-_l\right) + \bar d_2 \bm\sigma_\times \tau^-_\times \right] \delta(\bm r) \delta (\bm x - \bm r_k),
\end{eqnarray}
\end{widetext}
where $F_\pi = 92.4$ MeV is the pion decay constant, $m_\pi$ is the pion mass, $\bm r = \bm r_k - \bm r_l$, and $\hat{\bm r} \equiv \frac{\bm r}{r}$. The Yukawa functions are $Y_0(r) = \frac{e^{-m_\pi r}}{4\pi r}$ and $Y_2(r) = \frac{1}{m_\pi^2} r \frac{\partial}{\partial r} \frac{1}{r} \frac{\partial}{\partial r} Y_0(r)$. The compound spin and isospin operators are $\bm\sigma_\times = \bm\sigma_k \times \bm\sigma_l$ and $\tau^-_\times = (\tau_k \times \tau_l)^-$ \cite{Park_2003_PRC}. The dimensionless low-energy constants (LECs) are $\bar c_3, \bar c_4, \bar d_1, \bar d_2$ with the definition $\bar c_{3, 4} = m_N c_{3, 4}$, $\bar d_{1, 2} = \frac{m_N F_\pi^2}{g_A} d_{1, 2}$ \cite{Park_2003_PRC} and $\bar c_{D} \equiv \bar d_1 + 2 \bar d_2$.

\begin{figure*}[htbp]
  \centering
  \includegraphics[width=0.98\textwidth]{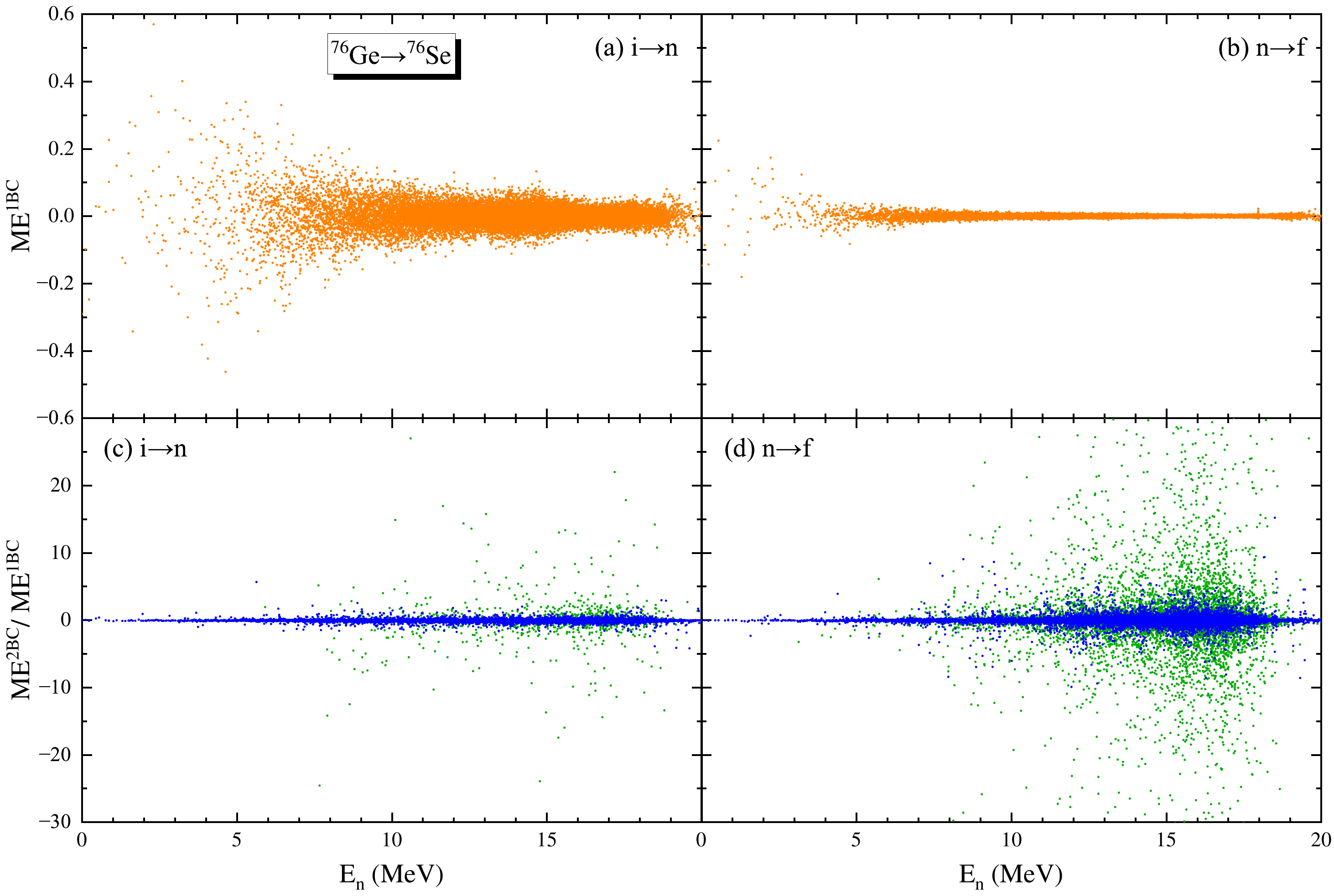}
  \caption{\label{fig.single_ME} (Color online) The single-$\beta$ matrix element (ME), i.e., $\int d \bm{x}_1 \langle 1_n^+ \| \hat{\mathcal{J}}(\bm{x}_1) \| 0_i^+ \rangle$ or $\int d \bm{x}_2 \langle 0_f^+ \| \hat{\mathcal{J}}(\bm{x}_2) \| 1_n^+ \rangle$ for the transitions from the initial ($i$) to the intermediate ($n$) levels, or the transitions from the intermediate to the final ($f$) levels, for the $^{76}$Ge $\rightarrow$ $^{76}$Se $2\nu\beta\beta$ decay as a function of $E_n$. Upper panels: the single-$\beta$ ME considering only the one-body current (1BC) , i.e., $\langle 1_n^+ \|  \bm\sigma \tau^- \| 0_i^+ \rangle$ or $\langle 0_f^+ \| \bm\sigma \tau^- \| 1_n^+ \rangle$. Lower panels: the ratio of the ME considering the two-body current (2BC) to the ME considering 1BC, where the green dots denote the cases that the absolute value of ME considering the 1BC is smaller than $0.001$.}
\end{figure*}

It is very challenging to consider both the 2BC term and the thousands of levels within each 1 MeV bin of the intermediate nucleus simultaneously, especially for open-shell (heavy deformed) nuclei. It is known from the Oslo experiments \cite{Guttormsen_Level_Density_EpJA_2015} and our recent calculations based on the projected shell model (PSM) \cite{JQWang_PRC_2023, JQWang_PRC_2025, Sun_EPJ_Web_2025} that the different orders of quasi-particle (qp) configurations are the basic building blocks for nuclear level density. To account for the high level density of the intermediate nucleus for $\beta\beta$ decays, we adopt the PSM for describing nuclear many-body wave functions, which starts from the many-body multi-qp configurations \cite{LJWang_2014_PRC_Rapid, BLWang_1stF_2024},
\begin{align} \label{eq.config}
  \textrm{ee}: \big\{ & |\Phi \rangle, 
               \hat{a}^\dag_{n_i} \hat{a}^\dag_{n_j} |\Phi \rangle,
               \hat{a}^\dag_{p_i} \hat{a}^\dag_{p_j} |\Phi \rangle,
               \hat{a}^\dag_{n_i} \hat{a}^\dag_{n_j} \hat{a}^\dag_{p_k} \hat{a}^\dag_{p_l} |\Phi \rangle, \nonumber\\ 
             & \hat{a}^\dag_{n_i} \hat{a}^\dag_{n_j} \hat{a}^\dag_{n_k} \hat{a}^\dag_{n_l} |\Phi \rangle, 
               \hat{a}^\dag_{p_i} \hat{a}^\dag_{p_j} \hat{a}^\dag_{p_k} \hat{a}^\dag_{p_l} |\Phi \rangle,   \big\} \nonumber \\
  \textrm{oo}: \big\{ & \hat{a}^\dag_{n_i} \hat{a}^\dag_{p_j}|\Phi \rangle, 
               \hat{a}^\dag_{n_i} \hat{a}^\dag_{n_j} \hat{a}^\dag_{n_k} \hat{a}^\dag_{p_l} |\Phi \rangle,
               \hat{a}^\dag_{n_i} \hat{a}^\dag_{p_j} \hat{a}^\dag_{p_k} \hat{a}^\dag_{p_l} |\Phi \rangle, \big\} 
\end{align}
for even-even (ee) and odd-odd (oo) nuclei respectively, where $|\Phi \rangle$ is the qp vacuum with associated intrinsic deformation and $\hat{a}^\dag_n (\hat{a}^\dag_p)$ labels the neutron (proton) qp creation operator. These multi-qp configurations are defined in the virtual intrinsic frame where some symmetries are broken \cite{Dobaczewski_PRR_2025}, which can be restored exactly by the projection technique. The rotational symmetry can be restored by the angular-momentum-projection operator $\hat{P}_{MK}^{J}$ \cite{Hara_PSM_review_1995},
\begin{eqnarray} \label{AMP_operator}
    \hat{P}^{J}_{MK} = \frac{2J + 1}{8\pi^2} \int d\Omega D^{J\ast}_{MK} (\Omega) \hat{R} (\Omega) ,
\end{eqnarray}
where $\hat{R}$ and $D_{MK}^{J}$ (with Euler angle $\Omega$) \cite{varshalovich1988quantum} are the rotation operator and Wigner $D$-function \cite{BLWang_2022_PRC} respectively. $K$ ($M$) is the  projection of the angular momentum $J$ in the intrinsic (laboratory) frame. From Eqs. (\ref{eq.config}, \ref{AMP_operator}) we can obtain the projected basis which is defined in the laboratory frame, the nuclear wave function can then be expanded in terms of the projected basis as, 
\begin{eqnarray} \label{eq.wave_function}
  | \Psi^{n}_{JM} \rangle = \sum_{K\kappa} f_{K\kappa}^{Jn} \hat{P}_{MK}^{J} | \Phi_{\kappa} \rangle ,
\end{eqnarray}
where $| \Phi_{\kappa} \rangle$ labels the multi-qp configurations in Eq. (\ref{eq.config}), and the coefficients $f$ can be obtained by solving the Hill-Wheeler equation with appropriate many-body Hamiltonian \cite{Hara_PSM_review_1995}. Since it is challenging to include the large configuration space in Eq. (\ref{eq.config}) and the 2BC in Eq. (\ref{eq.two_body_current}) simultaneously for heavy deformed nuclei, we adopt an effective Hamiltonian which includes the monopole and quadrupole terms in both the particle-hole and particle-particle channels as well as the two-body GT force \cite{BLWang_1stF_2024, LJWang_2018_PRC_GT}. In our PSM calculations, all the model parameters are adopted as in \cite{LJWang_2018_PRC_GT, HZhou_currents_arXiv} and three major shells with $N = 2,3,4$ are taken for neutrons and protons for the model space of $^{76}$Ge.

\section{\label{sec:result} Results and Discussion}

\begin{figure*}[htbp]
  \centering
  \includegraphics[width=0.98\textwidth]{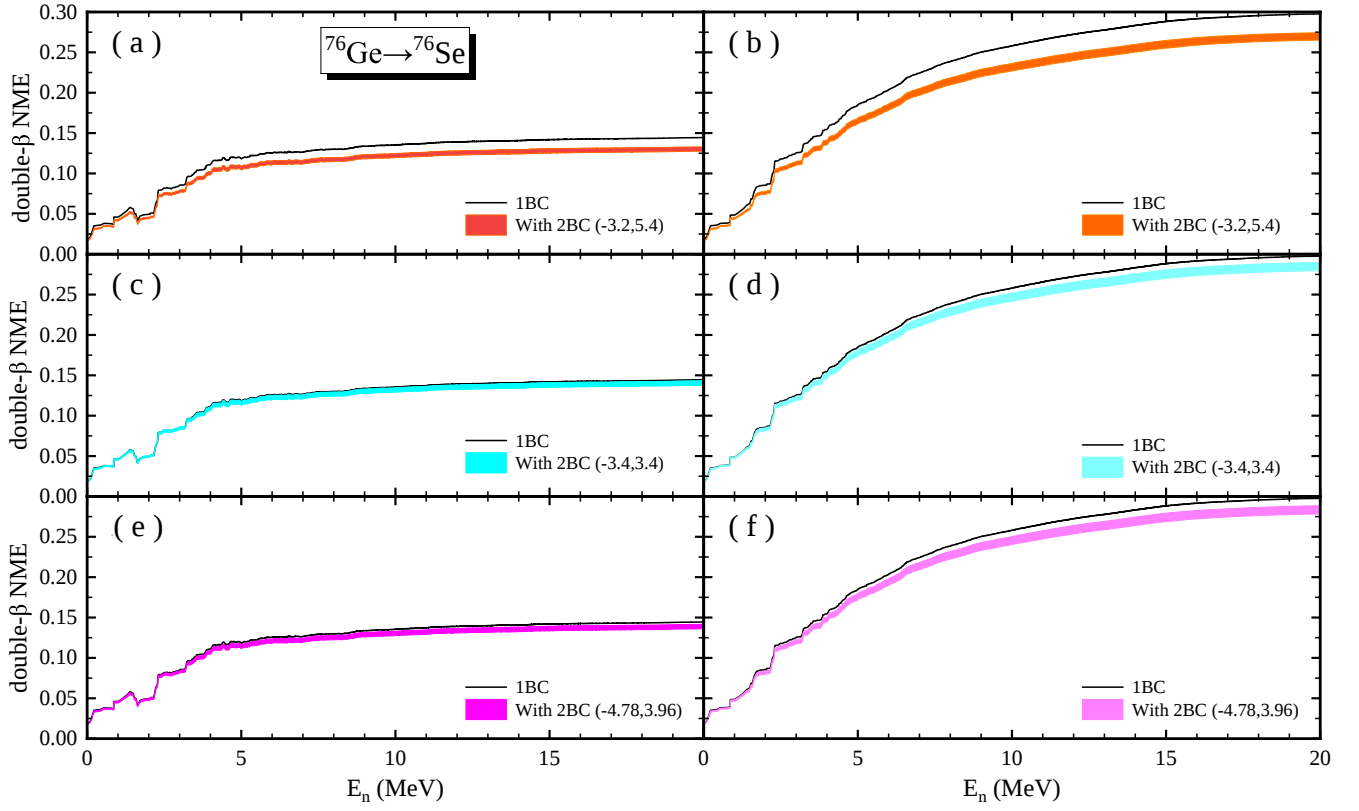}
  \caption{\label{fig.double_ME} (Color online) The cumulative $2\nu\beta\beta$-decay NME of $^{76}$Ge, considering the realistic single-$\beta$ matrix elements i.e., $m_e \sum^n_{k=1} \frac{M_k}{ E_k + E_0 }$ as shown in Eq.(\ref{eq.double_NME}) (left panels), or considering the absolute value of single-$\beta$ matrix elements i.e., $m_e \sum^n_{k=1} \frac{\left|M_k\right|}{ E_k + E_0 }$ (right panels). The calculations with only the 1BC are shown by black lines, while the calculations that further consider the 2BC are shown in colored bands. See the text for details. }
\end{figure*}

\begin{figure}[t] 
  \centering
  \includegraphics[width=0.49\textwidth]{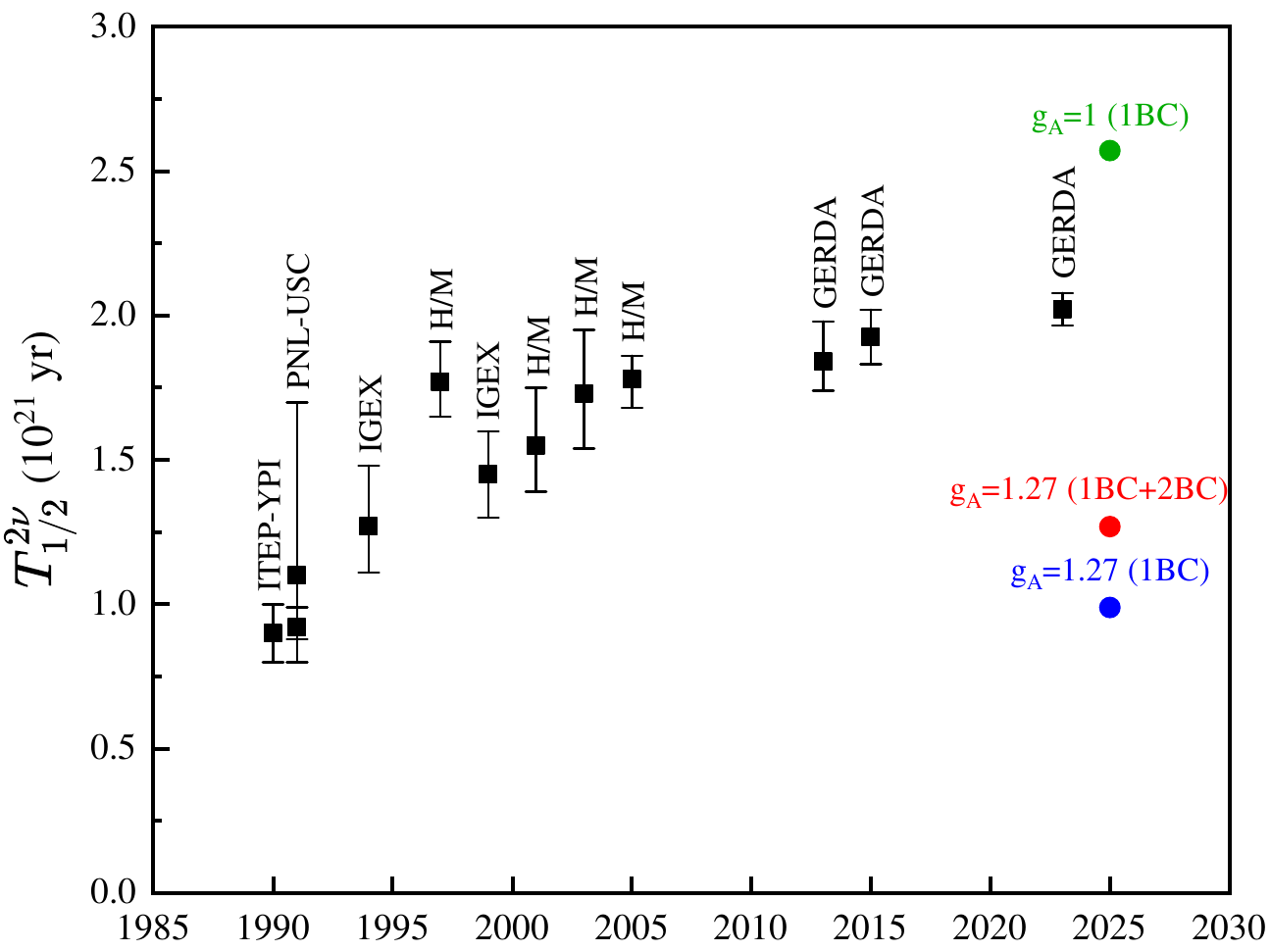}
  \caption{\label{fig.half_life} The calculated $2\nu\beta\beta$-decay half-life of $^{76}$Ge as compared with historical experimental measurements in different publication years \cite{Vasenko_MPLA_1990, Miley_1990_PRL, Avignone_1991_PLB, Avignone_1994_PPNP, Heidelberg_Moscow_1997_PRD, morales_1999_review, Dorr_NIM_2003, bakalyarov2005results, Agostini_2013_JPG, GERDA_2015_EpJC, Agostini_2023_PRL_final}. Calculations for cases with the bare $g_A=1.27$ or quenched $g_A=1.0$, and for cases considering only the 1BC or both 1BC and 2BC, are compared with each other. }
\end{figure}

To better analyze the $2\nu\beta\beta$-decay NME of $^{76}$Ge in Eq. (\ref{eq.double_NME}), we first show in Fig. \ref{fig.single_ME} the calculated single-$\beta$ matrix elements for the transitions from the initial $0^+_1$ level of the parent $^{76}$Ge to the intermediate $1^+_n$ levels of $^{76}$As (the $i \rightarrow n$ transitions), and the transitions from the intermediate $1^+_n$ levels of $^{76}$As to the final $0^+_1$ level of the daughter $^{76}$Se (the $n \rightarrow f$ transitions). To account for the high level density of the involved $1^+_n$ levels and keep the computational burden under control, in the PSM calculations, the configuration space (\ref{eq.config}) is truncated by $E_{qp} \lesssim 18$ MeV where $E_{qp}$ is the qp energy. In this way we obtained about $25, 000$ $1^+_n$ levels within $E_n \lesssim 18.0$ MeV and a few of levels beyond that. As seen from Fig. \ref{fig.single_ME} (a) and (b), on one hand, when only the 1BC is considered, the single-$\beta$ matrix elements of the $n \rightarrow f$ transitions are systematically much smaller than those of the $i \rightarrow n$ transitions. As seen from Eq. (\ref{eq.double_NME}), the very small single-$\beta$ matrix elements of the $n \rightarrow f$ transitions would lead to small $2\nu\beta\beta$-decay NME of $^{76}$Ge, which turns out to be as small as $\approx 0.1$ as seen from the black line in Fig. \ref{fig.double_ME}(a), and the corresponding half-life is then as long as $\approx 10^{21}$ years as seen from Fig. \ref{fig.half_life}. 

On the other hand, as seen from Fig. \ref{fig.single_ME} (a) and (b), with the increasing $E_n$, the single-$\beta$ matrix elements are large and sparse at low $E_n$ region but become very small and very dense at high $E_n$ region, especially for the $n \rightarrow f$ transitions. Specifically, for the $i \rightarrow n$ transitions ($n \rightarrow f$ transitions), after $E_n \gtrsim 9$ MeV ($E_n \gtrsim 5$ MeV) the single-$\beta$ matrix elements become very small and very dense, and exhibit seemingly random sign patterns. As seen from Eq. (\ref{eq.double_NME}), the $2\nu\beta\beta$-decay NME is a sum over the product of the single-$\beta$ matrix elements for the $i \rightarrow n$ and $n \rightarrow f$ transitions. Therefore, the different signs between the single-$\beta$ matrix elements of the $i \rightarrow n$ and $n \rightarrow f$ transitions may cause cancellations in deriving the $2\nu\beta\beta$-decay NME, and the highly fragmented small values and seemingly random-sign patterns at high $E_n$ region will probably lead to total cancellation in the summation. To illustrate this, we show in Fig. \ref{fig.double_ME}(a) the cumulative $2\nu\beta\beta$-decay NME of $^{76}$Ge where the black line labels the case of considering only the 1BC, i.e., $m_e \sum^n_{k=1} \frac{ \langle 0_f^+ \| \bm\sigma\tau^- \| 1^+_k \rangle \langle 1^+_k \| \bm\sigma\tau^- \| 0^+_i \rangle }{ E_k + E_0 }$. It is seen that the cumulative $2\nu\beta\beta$-decay NME sometimes increases and sometimes decreases with the increasing $E_n$, and almost saturates and converges at $E_n \approx 5$ MeV, due to the cancellation mechanism discussed above.

The cancellation phenomenon is significant in several respects. On one hand, it indicates that the sign properties of the single-$\beta$ matrix elements play a crucial role in deriving the corresponding $2\nu\beta\beta$-decay NME. The signs of the single-$\beta$ matrix elements are not observables, but they are important for the $\beta\beta$-decay NME. Because the $0\nu\beta\beta$-decay NME is not an observable and both the $2\nu\beta\beta$ and $0\nu\beta\beta$ decay modes share the identical initial and final nuclear states, the $2\nu\beta\beta$ decay can then provide helpful constraints on nuclear-structure models that are applied to calculate the $0\nu\beta\beta$-decay NME. An important way for the constraint is detecting the corresponding GT transition strength, $B(\text{GT})$, of the single-$\beta$ transitions by for example, the charge-exchange reactions \cite{Thies_76Ge_PRC_2012}. However, the $B(\text{GT})$ only reflects the absolute value of single-$\beta$ matrix elements and their sign properties get lost. To illustrate this, the cumulative $2\nu\beta\beta$-decay NME neglecting the sign properties of the single-$\beta$ matrix elements is shown in Fig. \ref{fig.double_ME}(b), where the black line labels the case of considering only the 1BC, i.e., $m_e \sum^n_{k=1} \frac{ \left| \langle 0_f^+ \| \bm\sigma\tau^- \| 1^+_k \rangle \langle 1^+_k \| \bm\sigma\tau^- \| 0^+_i \rangle \right| }{ E_k + E_0 }$. It is seen from Fig. \ref{fig.double_ME}(b) that, since the sign properties get lost, such a hypothetical cumulative $2\nu\beta\beta$-decay NME keeps increasing with the increase of $E_n$, and is expected to increase more rapidly in the high $E_n$ region if we relax the truncation on the configuration space (\ref{eq.config}). By comparison, the realistic cumulative $2\nu\beta\beta$-decay NME shown in Fig. \ref{fig.double_ME}(a) almost saturates at $E_n \approx 5$ MeV. On the other hand, the calculations shown in Fig. \ref{fig.double_ME}(a) indicate that for the $2\nu\beta\beta$-decay NME, only the single-$\beta$ transitions involving the low-lying levels of the intermediate nucleus with low $E_n$ contribute effectively. This is important for constraining and studying the $2\nu\beta\beta$-decay NME by the charge-exchange reactions and model calculations, since both of them are difficult to treat nuclear high-lying states with very high level density. 

It is interesting that the cancellation phenomenon in $2\nu\beta\beta$-decay NME was previously studied by different models including the shell model \cite{Brown_PRC_1990, Nakada_NPA_1996} and the quasiparticle random-phase approximation (QRPA) \cite{Suhonen_PRC_1997, Raduta_PRC_2005, Fang_PRC_2010}. For the light doubly-magic nucleus $^{48}$Ca, the shell model calculations with limited model space found that with the increase of $E_n$, the $2\nu\beta\beta$-decay NME increases and then saturates at $E_n \approx 10$ MeV, besides, the saturation behavior is very sensitive to the truncation of model space \cite{Brown_PRC_1990}. The QRPA calculations found that the $2\nu\beta\beta$-decay NME of $^{48}$Ca decreases with $E_n$ and saturates when $E_n \gtrsim 15$ MeV \cite{Raduta_PRC_2005}. For the open-shell (medium-) heavy deformed nucleus $^{76}$Ge, the $2\nu\beta\beta$-decay NME is found to saturate when $E_n \gtrsim 10$ MeV from the QRPA calculations \cite{Fang_PRC_2010}. As shown in Fig. \ref{fig.double_ME}(a), our PSM calculations with large model space and large configuration space indicate that the $2\nu\beta\beta$-decay NME saturates at a lower energy $E_n \approx 5$ MeV.


Finally we discuss the effect of many-body currents. In Fig. \ref{fig.single_ME}(c) and (d) we show the ratio of the single-$\beta$ matrix element considering the 2BC to the single-$\beta$ matrix element considering the 1BC, where the coupling constants in the 2BC are adopted as $(c_3, c_4) = (-3.2, 5.4)$ \cite{coupling_PRC_2003_R} and $\bar c_D = -2.0$. It is seen that at the low $E_n$ region with $E_n \lesssim 5$ MeV, the contribution of 2BC to the single-$\beta$ matrix elements is small. At the high $E_n$ region with $E_n \gtrsim 8$ MeV, although the contribution of the 2BC is much larger than that of 1BC for many transitions, in most cases the corresponding 1BC matrix elements are too small (see the green dots) and the signs of the ratio are seemingly random. These features imply that the contribution of 2BC to the $2\nu\beta\beta$-decay NME will not be significant. In Fig. \ref{fig.double_ME} the effects of the 2BC on the cumulative $2\nu\beta\beta$-decay NME are shown by different colored bands, which correspond to adopting different values for the coupling constants, i.e., $(c_3, c_4) =$ $(-3.2, 5.4)$ \cite{coupling_PRC_2003_R}, $(-3.4, 3.4)$ \cite{coupling_NPA_2005}, $(-4.78, 3.96)$ \cite{coupling_PRC_2003}, and the width of the colored error band comes from taking the range of $\bar c_D$ from $-2.0$ to $2.0$. It is seen from Fig. \ref{fig.double_ME}(a), (c) and (e) that the 2BC provides about $10\%$ quenching to the $2\nu\beta\beta$-decay NME of $^{76}$Ge, while the effect of the 2BC is sensitive to the employed coupling constants where smaller $\bar c_D$ leads to larger quenching. The largest effect comes from the couplings with $(c_3, c_4) =$ $(-3.2, 5.4)$ and $\bar c_D = -0.2$, the corresponding NME is quenched from 0.144 to 0.127 by about $12\%$, as shown in Fig. \ref{fig.double_ME}(a). Accordingly, the calculated $2\nu\beta\beta$-decay half-life is increased from $1.0 \times 10^{21}$ yr to $1.3 \times 10^{21}$ yr, as shown by blue and red dots in Fig. \ref{fig.half_life}. By comparison, phenomenologically quenching $g_A = 1.0$ (as in the literature \cite{Satula_PRC_2025_Editor}) yields a half-life of about $2.5 \times 10^{21}$ yr as seen from Fig. \ref{fig.half_life}. These findings indicate that the corresponding coupling constants in the 2BC need to be reliably determined.

\section{\label{sec:summary} Summary and outlook }

In summary, we provide a microscopic method to calculate the $2\nu\beta\beta$-decay NME for any open-shell heavy (deformed) nuclei where both the nuclear high level density and the two-body current can be treated properly. We found that the $2\nu\beta\beta$-decay NME of $^{76}$Ge saturates when only transitions involving low-lying states of the intermediate nucleus are considered, because the products of different single-$\beta$ matrix elements cancel each other upon summation. The two-body current is found to provide about $10\%$ quenching to the $2\nu\beta\beta$-decay NME of $^{76}$Ge and about $30\%$ enhancement to the corresponding half-life, where the effect of the two-body current is sensitive to the corresponding coupling constants. Similar work on the more important counterpart of the $0\nu\beta\beta$-decay NME will be completed in the near future.

\begin{acknowledgments}
  This work is supported by the National Natural Science Foundation of China (Grant No. 12275225) and by the New Chongqing Youth Innovation Talent Project (Grant No. CSTB2025YITP-QCRCX0055).  
\end{acknowledgments}


%

\end{document}